\documentclass[fleqn,twoside]{article}
\usepackage{fleqn,espcrc2,epsf}

\usepackage{graphicx}

\newcommand{\be}{\begin{equation}}
\newcommand{\ee}{\end{equation}}
\newcommand{\AmS}{{\protect\the\textfont2
  A\kern-.1667em\lower.5ex\hbox{M}\kern-.125emS}}

\newcommand{\nn}{\nonumber}

\title{Improving Dynamical Domain-Wall Fermion Simulations
\thanks{This work was conducted on the QCDSP machines at Columbia
     University and the RIKEN-BNL Research Center. LL and RM are
     supported by the US DOE.
}}

\author{L.~Levkova\address{Department of Physics, Columbia 
University, New York, NY, 10027} and R.~Mawhinney$^{\rm a}$}
\begin{document}
\bibliographystyle{apsrev}

\begin{abstract}
We report on studies of the chiral properties of dynamical domain
  wall fermions combined with the DBW2 gauge action for different gauge
  couplings and fermion masses.  For quenched theories, the DBW2 action
  gives a residual chiral symmetry breaking much smaller than what was
  found with more traditional choices for the gauge action.  Our goal
  is to investigate the possibilities which this and further improvements
  provide for the study of QCD thermodynamics and other simulations at
  stronger couplings.
\vspace{1pc}
\end{abstract}

\maketitle

\section{INTRODUCTION}

For the lattice domain-wall fermion formulation there are two independent approaches which can 
be used to reduce the residual chiral symmetry breaking ($m_{\rm res}$) for practical sizes of $L_s$.
One way is to modify the fermion part of the action
and another way is to improve the gauge action used. 
In this paper we take the second approach by studying the DBW2 gauge action
in dynamical DWF calculations and considering the effects of also adding 
an adjoint term to the gauge action.

The DBW2 gauge action reduces $m_{\rm res}$ at finite
$L_s$ in quenched DWF simulations by suppressing the 
topological lattice dislocations in the gauge field configurations\cite{inst,rbc}.
These simulations\cite{rbc} find that $m_{\rm res}$ is
about two orders of magnitude smaller than the corresponding value for the Wilson action at a given lattice scale.
Here we investigate the size of that improvement for the corresponding
dynamical DWF cases at zero and finite temperatures.

The form of the DBW2 action is as follows:
\begin{small}
\[
S_{DBW2} = -\frac{\beta}{N_c}\left((1-8c_1)
     \sum_{x;\mu<\nu} {\rm Re} {\rm Tr} P_{\mu\nu}(x)\right.\]
\[\hspace{1.5cm} + \left. c_1 \sum_{x;\mu \neq \nu } {\rm Re} {\rm Tr}R_{\mu\nu}(x)\right).\]  
\end{small}
It is a renormalization group improved action where $c_1=-1.4069$ is nonperturbatively computed\cite{dbw2}.

We  extend the study of lattice dislocation suppression  by adding 
an adjoint term to the DBW2 gauge action for the dynamical DWF case in the 
hope that it will improve further $m_{\rm res}$
in the strong coupling sector without affecting
the long distance physics much.
\vspace{-0.2cm}
\section{RESIDUAL MASS FOR DYNAMICAL ZERO-TEMPERATURE SIMULATIONS}
\vspace{-0.15cm}
As a measure of the residual chiral symmetry breaking we use the ratio\cite{rbc}:
\begin{small}
\[
m_{\rm res}=\left.\frac{\sum_{x,y}\langle J_{5q}^a(y,t)J_5^a(x,0)\rangle}{\sum_{x,y}
\langle J_{5}^a(y,t)J_5^a(x,0)\rangle}\right|_{t\geq t_{min}} .\nn
\]
\end{small}
%
%

Table~1 contains the parameters of the dynamical zero-temperature runs and the respective values for $m_{\rm res}$.
The main feature of these results is that $m_{\rm res}$ significantly increases with 
the gauge coupling. For the weak coupling
of $\beta=0.80$, $m_{\rm res}$ is quite small and appears independent of the quark mass. 
For the strongest coupling from Table~1, $\beta=0.70$, $m_{\rm res}$ has 
risen to as much as one third of the input quark mass.
\begin{table}[ht]
\begin{small}
\begin{tabular}{lllll}
\hline \hline
$\beta$ & $m_f$  & $m_\rho$, $\chi^2/Dof$  & $m_{\rm res}$ & \# conf.\\
\hline \hline
0.70 	& 0.026 & 0.830(5), 1.4 & $0.0094(1)$&   32          \\

0.75   & 0.022 &  0.667(7), 0.8    & $0.00405(7)$ &  41     \\

0.80  & 0.020 &  0.539(6),  1.6       & $0.00137(2)$ &  75     \\

0.80  & 0.030 & 0.598(8),  0.8        & $0.00136(2)$& 72    \\

0.80  & 0.040  & 0.645(8), 1.7      & $0.00137(3)$ & 32  \\
\hline \hline
\end{tabular}
\caption{\small{Parameters and $m_{\rm res}$  of dynamical DWF zero-temperature calculations.
 All runs have volume $16^3\times 32$, $L_s=12$ and $M_5=1.8$.}}
\end{small}
\vspace{-0.5cm}
\end{table}

\section{RESIDUAL MASS FOR DYNAMICAL FINITE TEMPERATURE SIMULATIONS}
For the finite temperature dynamical case the run parameters are shown in Table~2.
We measure $m_{\rm res}$ from screening pseudoscalar correlators in the spatial z direction. Figure~1 shows the ratio
$m_{\rm res}$ as a function of the spatial coordinate z. Ideally after some z, $m_{\rm res}$ should be virtually 
a constant and the corresponding graph should have a plateau. 
In the case of $N_t = 4$ and $\beta = 0.75$ the lack of
a plateau in $m_{\rm res}$ (which rises noticeably with
increasing $z$) likely indicates that the desired
propagation in the four-dimensional boundaries is modified
by propagation into the bulk fifth dimension.  This effect
is noticeably reduced at the weaker coupling of $\beta =
0.80$ and is the subject of further investigation.
%
\begin{table}[ht]
\vspace{-0.4cm}
\begin{small}
\begin{tabular}{llllll}
\hline \hline
$\beta$ & $m_f$  & $N_t$ & $m_\rho$& $m_\pi$   & \# conf.\\
\hline \hline
0.75 	& 0.022 &  4 	&  1.465(3) &1.344(4) &  74    \\

0.80   & 0.020 &  4      & 1.472(3)&1.379(4)   &  37     \\

0.80  & 0.020 &   8    &  0.737(6)& 0.53(1)  &    34     \\
\hline \hline
\end{tabular}
\caption{\small{Parameters of dynamical DWF finite temperature calculations. All runs have spatial volume $16^3$, $L_s=12$ and $M_5=1.8.$}}
\end{small}
\vspace{-0.5cm}
\end {table}
\begin{figure}[ht]
\epsfxsize=\hsize
\epsfbox{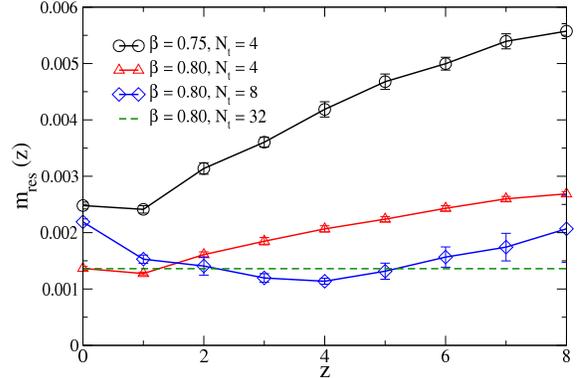}
\vspace{-1cm}
\caption{\small{$m_{\rm res}(z)$ for $N_t=4$ and 8.}}
\end{figure}
\section{PROPERTIES OF THE DYNAMICAL GAUGE CONFIGURATIONS}
From the results of section 2 and 3 it is clear that the gauge coupling has 
a strong effect on $m_{\rm res}$. We want to investigate the properties of
the dynamical DBW2 gauge configurations which 
affect directly the residual chiral symmetry breaking. One way of doing that is by 
studying the spectral flow of the hermitian wilson dirac operator $\gamma_5 D_W$. Figure~2 shows some typical spectral flows for three
of the zero-temperature dynamical runs from Table~1. The two properties of the spectral flow, gap width and number
 of crossings, which contribute to $m_{\rm res}$, are changing dramatically with $\beta$. 
For the weak coupling
 of $\beta=0.80$ the gap is open, there are few crossings and consequently
$m_{\rm res}$ is small. The situation for the stronger couplings worsens to the point that at the
 strongest coupling of  $\beta=0.70$ the gap is virtually closed and $m_{\rm res}$ is about an
 order of magnitude larger.

Another way to study the gauge configurations is to plot directly their plaquette distributions.
Figure~3 shows the comparison of the plaquette distributions of the dynamical runs with
the quenched DBW2 case. The conclusion is that the fermion determinant has the effect 
of pulling the distribution towards more negative plaquette values, which are associated with lattice 
dislocations and larger $m_{\rm res}$.
\begin{figure}[ht]
\vspace{-0.7cm}
\epsfxsize=\hsize
\begin{center}
\epsfbox{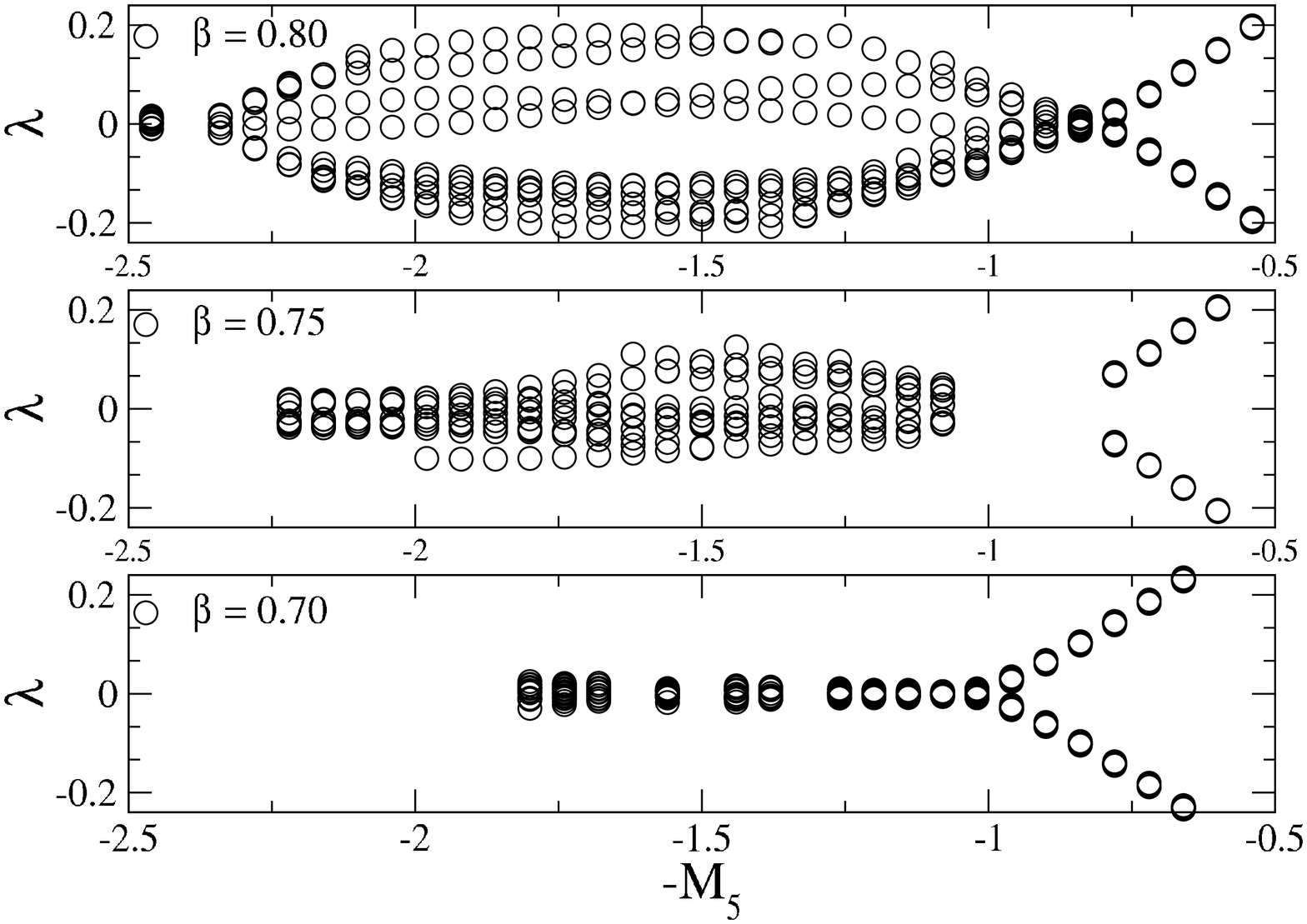}
\end{center}
\vspace{-1.5cm}
\caption{\small{Spectral flows of $\gamma_5 D_W$ for three dynamical zero-temperature runs from Table~1.}}
\vspace{-0.5cm}
\end{figure}
\begin{figure}[ht]
\vspace{-0.7cm}
\epsfxsize=\hsize
\begin{center}
\epsfbox{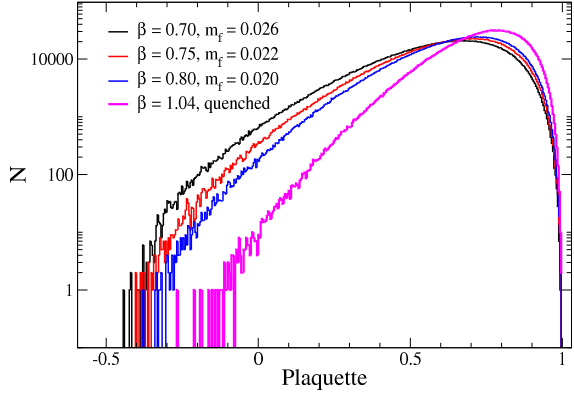}
\end{center}
\vspace{-1.5cm}
\caption{\small{Plaquette distributions of dynamical runs from Table~1 compared with the quenched DBW2 case.}}
\vspace{-0.6cm}
\end{figure}
\section{ADDING AN ADJOINT TERM TO DBW2}
Our idea to decrease $m_{\rm res}$ in the strong coupling sector is to modify
the DBW2 action by adding an adjoint term to it in the way shown below:
\begin{small}
\begin{eqnarray}
\label{eq:asym_gauge_action}
S_{\small \rm  dbw2+adj} 
  &=& -\frac{\beta}{N_c}\left((1-8c_1)\left[c_f
     \sum_{x;\mu<\nu} {\rm Re} {\rm Tr} P_{\mu\nu}(x)\right.\right. \nonumber\\
 &&\hspace{-1.8cm}+ \left.\left.\frac{c_a}{N_c}\sum_{x;\mu<\nu}\left|{\rm Tr} P_{\mu\nu}(x)\right|^2 \right]
 +  c_1 \sum_{x;\mu \neq \nu } {\rm Re} {\rm Tr}R_{\mu\nu}(x)\right)
     \nonumber,
\end{eqnarray}
\end{small}
\hspace{-0.14cm}where we choose $c_f+2c_a=1$, to keep the normalization
for this three term action the same as for the plaquette action
in the continuum limit.
We expect that the modified action will suppress more strongly the negative tails of the plaquette
distributions and thus it will get rid of the lattice dislocations which increase $m_{\rm res}$. 

Table~3 summarizes the results for $m_{\rm res}$ for three runs with the modified DBW2 action. The values for 
$m_{\rm res}$ are an order of magnitude smaller than the values for the unmodified DBW2 with the same parameters. 
However to conclude that there is an improvement at a given lattice 
scale we need to measure the lattice spacings for the runs from Table~3.
All of those runs turned out to be in the deconfined phase and to measure the
scales correctly we need to repeat the calculation at a larger volume.
Nevertheless, Figure~4 shows the effect of the adjoint term on the plaquette distributions by
comparing runs from Table~3 with runs with the original DBW2 which have the closest in shape plaquette distribution,
although the parameters are different. The modified DBW2 renders the negative distribution tails shorter by comparison.
If similar plaquette distributions correspond to similar scales then there might
be an improvement of roughly two times smaller $m_{\rm res}$ for the action with the adjoint term.
%
\begin{table}[ht]
\vspace{-1.cm}
\begin{small}
\begin{tabular}{llllll}
\hline \hline
$\beta$ & $m_f$  & $c_f$ & $c_a$  & $m_{\rm res}$ & \# conf.\\
\hline \hline
0.65   &  0.026 & 1.2   & -0.1 & $0.0031(1)$ & 168\\

0.70 	& 0.022 &1.2  & -0.1 &$0.00083(7)$&   36          \\

0.75   & 0.022 & 1.2   & -0.1 & $0.00045(3)$ &  59     \\
\hline \hline
\end{tabular}
\end{small}
\caption{\small{$m_{\rm res}$ results for dynamical DBW2 with adjoint term runs
on $8^3\times 16$ volume, $L_s=12$ and $M_5=1.8.$}}
\vspace{-1.2cm}
\end {table}
\begin{figure}
\epsfxsize=\hsize
\begin{center}
\epsfbox{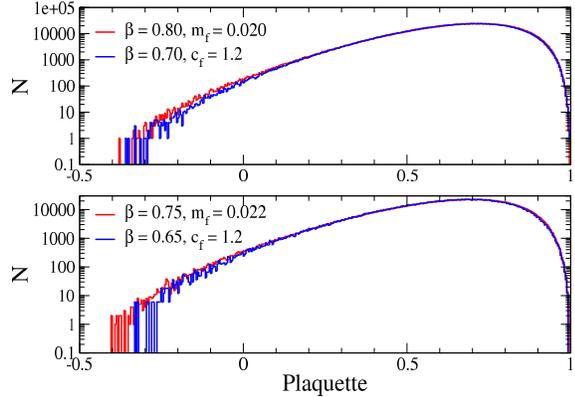}
\end{center}
\vspace{-1.5cm}
\caption{\small{Plaquette distribution for two of the runs from Table~3 compared with the unmodified 
DBW2 case with the closest in shape distribution. }}
\vspace{-0.5cm}
\end{figure}
\section{CONCLUSIONS}
The DBW2 action significantly suppresses lattice dislocations and reduces $m_{\rm res}$ at zero 
temperature for dynamical DWF runs at weak couplings.
For $N_t=4$ there is an unphysical increase in $m_{\rm res}$. For $N_t=8$,
 plots of $m_{\rm res}$ are likely showing a plateau at weak coupling.
The addition of an adjoint term to the DBW2 action in order to reduce further 
$m_{\rm res}$ is still under investigation, but it shows promise for improvement.

\end{document}